# Linewidth broadening of a quantum dot coupled to an off-resonant cavity

Arka Majumdar,[*] Andrei Faraon,[†] Erik D. Kim, Dirk Englund,[‡] and Jelena Vučković

*E.L.Ginzton Laboratory,*
*Stanford University, Stanford, CA, 94305*

Hyochul Kim and Pierre Petroff

*Materials Department, University of California,*
*Santa Barbara, CA 93106*

We study the coupling between a photonic crystal cavity and an off-resonant quantum dot under resonant excitation of the cavity or the quantum dot. Linewidths of the quantum dot and the cavity as a function of the excitation laser power are measured. We show that the linewidth of the quantum dot, measured by observing the cavity emission, is significantly broadened compared to the theoretical estimate. This indicates additional incoherent coupling between the quantum dot and the cavity.

Recent demonstrations of cavity quantum electrodynamics (CQED) with a single quantum dot (QD) coupled to a semiconductor micro-cavity show the great potential of this system for developing robust, scalable quantum information processing devices [1, 2, 3, 4]. However, unlike ultra-cold atoms, QDs constantly interact with their local environments and this interaction plays a significant role in CQED experiments with QDs. For example, several experiments have reported the observation of cavity emission even when the QD is far detuned ($\sim 3-10$ meV) from the cavity resonance, in contrast with atomic CQED experiments. This unexpected non-resonant QD-cavity coupling is observed both in photoluminescence, where the QD is excited by creating carriers above the band-gap of the GaAs surrounding the QD [2, 5, 6] and in the cavity luminescence under resonant excitation of the QD [7, 8]. Recent theoretical investigations have attributed the off-resonant coupling to several different causes including pure dephasing [9], phonon relaxation [10], multi-exciton complexes [11] and charges surrounding the QD [12].

In this paper, we experimentally study the process responsible for transferring photons between the QD and off-resonant cavity mode, under resonant excitation of the QD or the cavity. We derive an analytical expression for the QD linewidth based on pure dephasing and coupling to the cavity, but find that experimentally obtained linewidths are larger than that predicted by the theory. We attribute this to an additional incoherent coupling mechanism between the QD and the cavity.

When an off-resonant QD that is coupled to a cavity is coherently driven by a laser field, the QD is dressed by both the cavity and the laser field. In the absence of a driving laser, the dynamics of a coupled QD-cavity system is described by the Jaynes-Cummings Hamiltonian

$$H_{JC} = \hbar\omega_c a^\dagger a + \hbar\omega_d \sigma^\dagger \sigma + \hbar g(\sigma^\dagger a + \sigma a^\dagger) \quad (1)$$

Here, $\omega_c$ and $\omega_d$ are the cavity and the QD resonance frequency, respectively, $\sigma$ is the lowering operator for the QD, $a$ is the annihilation operator for the cavity photon and $g$ is the coherent interaction strength between the QD and the cavity. The eigen-frequencies $\omega_\pm$ of the coupled system are given by [1]

$$\omega_\pm = \frac{\omega_c + \omega_d}{2} - i\frac{\kappa + \gamma}{2} \pm \sqrt{g^2 + \frac{1}{4}(\delta - i(\kappa - \gamma)^2)} \quad (2)$$

where $2\kappa$ and $2\gamma$ are the cavity energy decay rate and the QD spontaneous emission rate, respectively and $\delta$ is the QD-cavity detuning $\omega_d - \omega_c$. When the coherent interaction strength $g$ is greater than the decay rates $\kappa$ and $\gamma$, the system is in strong coupling regime, and the eigen-states of $H_{JC}$ are polaritons possessing the characteristics of both the cavity and the QD. In this regime, when the QD-cavity detuning $\delta = 0$, the linewidth of the polaritons is $\kappa + \gamma$. However, when the QD-cavity detuning $\delta$ is much greater than $g$, the system is in the dispersive CQED regime. In this regime, one polariton develops a cavity-like character while the other becomes more QD-like. The linewidths $\Gamma_c$ and $\Gamma_{qd}$ of the cavity-like and QD-like polaritons, respectively, are given by (with a pure QD dephasing rate of $\gamma_d$) [13]

$$\Gamma_c \simeq 2\kappa + 2\left(\frac{g}{\delta}\right)^2 \gamma \quad (3)$$

$$\Gamma_{qd} \simeq 2(\gamma + \gamma_d) + 2\left(\frac{g}{\delta}\right)^2 \kappa \quad (4)$$

The linewidth $\Gamma_{qd}$ can be interpreted as a combination of the QD spontaneous emission rate ($2\gamma$) and the QD emission rate into the cavity mode $2(g/\delta)^2 \kappa$.

---
[*]Electronic address: arkam@stanford.edu
[†]Currently at H.P Laboratories, Palo Alto, CA-94304
[‡]Currently at Columbia University, New York City, NY

On the other hand, when a QD is coherently driven by a laser field in the absence of any optical cavity, the system dynamics is described by the Master equation

$$\frac{d\rho}{dt} = -\frac{i}{\hbar}[H,\rho] + 2\gamma\mathcal{L}[\sigma] + \frac{\gamma_d}{2}\mathcal{L}[\sigma^\dagger\sigma] \quad (5)$$

Here $\rho$ is the density matrix of the QD optical transition and $\gamma_d$ is the pure dephasing rate. $\mathcal{L}[D]$ is the Lindblad operator for an operator $D$ and is given by

$$\mathcal{L}[D] = D\rho D^\dagger - \frac{1}{2}D^\dagger D\rho - \frac{1}{2}\rho D^\dagger D \quad (6)$$

The Hamiltonian $H$ describing the coherent dynamics of the driven QD is given by

$$H = \hbar\omega_d \sigma^\dagger \sigma + \hbar\frac{\Omega}{2}(\sigma e^{-i\omega_l t} + \sigma^\dagger e^{i\omega_l t}) \quad (7)$$

where, $\omega_d$ and $\omega_l$ are the QD resonance and the driving laser frequency, respectively, and $\Omega$ is the Rabi frequency of the driving laser field. In solving the Master equation (Eq. 5), it is found that the intensity $I$ of the QD resonance fluorescence for $\omega_d = \omega_l$ is given by

$$I = \frac{\frac{\Omega^2}{4\gamma(\gamma+\gamma_d)}}{1 + \frac{\Omega^2}{2\gamma(\gamma+\gamma_d)}} \propto \frac{\tilde{P}}{1+\tilde{P}} \quad (8)$$

where $\tilde{P} = \frac{\Omega^2}{2\gamma(\gamma+\gamma_d)}$. The QD linewidth $\Delta\omega$ is given by

$$\Delta\omega = 2(\gamma+\gamma_d)\sqrt{1 + \frac{\Omega^2}{2\gamma(\gamma+\gamma_d)}} \propto \sqrt{1+\tilde{P}} \quad (9)$$

The broadening of the QD linewidth with laser excitation power occurs due to increasing stimulated emission caused by the laser field and is known as power broadening. Such power broadening of the QD linewidth has been reported by several other groups [14, 15].

Following the discussion above, the linewidth $\Delta\omega$ of a resonantly driven QD that is coupled to an off-resonant cavity has contributions from both the increased emission rate in the cavity mode and the increasing stimulated emission due to the driving laser. As the QD is detuned from the cavity (and hence the laser driving the QD resonantly is also detuned from the cavity), the QD emission into cavity mode and the stimulated emission into the driving laser mode are independent and $\Delta\omega$ is given by

$$\Delta\omega = 2\left(\frac{g}{\delta}\right)^2 \kappa + 2(\gamma+\gamma_d)\sqrt{1+\tilde{P}} \\ = \Delta\omega_c + \Delta\omega_0 \sqrt{1+\tilde{P}} \quad (10)$$

Here, $\Delta\omega_c = 2(g/\delta)^2 \kappa$ and $\Delta\omega_0 = 2(\gamma+\gamma_d)$. Similarly, as the cavity is coupled to the QD, the cavity-like polariton linewidth contains a contribution from the QD emission, as evident from Eq. 3. However as the cavity loss rate $2\kappa$ is much greater than the QD spontaneous emission rate $2\gamma$, the modification of the cavity linewidth is negligible. From now on, we will refer to the cavity-like polariton as the "cavity" and QD-like polariton as the "QD".

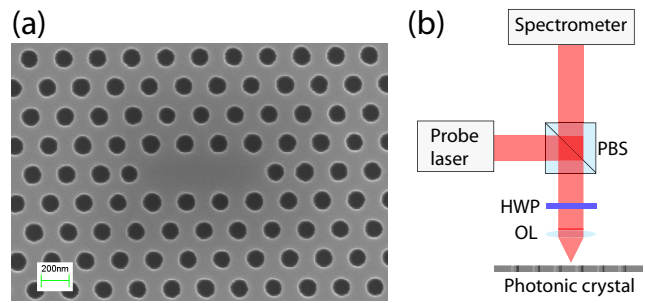

FIG. 1: (color online)(a) Scanning electron micrograph of the fabricated photonic crystal cavity. (b) The experimental setup. An objective lens (OL) with a numerical aperture of 0.75 is used in front of the cryostat to image the chip. A half wave plate (HWP) is used to adjust the excitation polarization relative to the cavity axis. A polarizing beam splitter (PBS) is used to perform cross-polarized reflectivity measurements. Details of the experimental setup are given in [1].

Experiments are performed in a helium-flow cryostat at cryogenic temperatures ($\sim 30-55$ K) on self-assembled InAs QDs embedded in a GaAs photonic crystal cavity [1]. The 160nm GaAs membrane used to fabricate the photonic crystal is grown by molecular beam epitaxy on top of a GaAs (100) wafer. The GaAs membrane sits on a 918 nm sacrificial layer of $Al_{0.8}Ga_{0.2}As$. Under the sacrificial layer, a 10-period distributed Bragg reflector, consisting of a quarter-wave AlAs/GaAs stack, is used to increase the collection into the objective lens. The photonic crystal was fabricated using electron beam lithography, dry plasma etching, and wet etching of the sacrificial layer in hydrofluoric acid (6%). A scanning electron micrograph of a photonic crystal cavity along with a diagram of the experimental setup is shown in Fig. 1.

We perform two different types of experiments to study the off-resonant QD-cavity coupling. For the first type, a narrow bandwidth ($\sim 300$ kHz) laser is scanned across the QD optical transition while the emission at the cavity wavelength is observed. In the second type, the laser is scanned across the cavity linewidth and the QD emission is observed. Figs. 2 (a), (b) show the cavity and QD emission spectra for the first and second experiments, respectively. Figs. 2 (c), (d) show the integrated cavity and QD intensities as we scan the laser across the QD and the cavity, respectively. Lorentzian fits to the cavity and the QD intensities as a function of laser wavelength enable estimation of the QD and the cavity linewidths, respectively.

The first type of experiment is performed on three different QD-cavity systems for different detunings between the cavity and the QD transition. Details of three systems are given in the Table I. The detuning between the cavity and a particular QD transition is controlled by varying the sample temperature. As the limited temperature tuning range limits the range of achievable QD-cavity detunings, multiple QDs must be chosen to cover an extended range of detunings. However, all three systems show similar qualitative behavior.

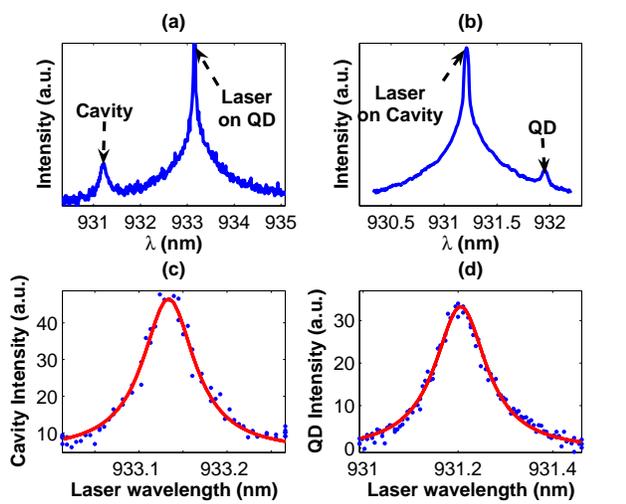

FIG. 2: (color online)(a) Cavity emission when the QD is resonantly excited. (b) QD emission when the cavity is resonantly excited. Experiments to obtain (a) and (b) are performed at 55 K. The cavity wavelength is 931.2 nm. The QD resonances are at (a) 933.15 nm and (b) 931.9 nm. Emission from the QD at 933.15 nm is very weak under resonant excitation of the cavity. Hence, for (b) another QD at 931.9 nm is used. (c) Integrated cavity emission as a function of the pump laser wavelength when the QD is resonantly excited [as in (a)]. The solid line is a Lorentzian fit (with a linewidth of 0.0879 nm). (d) Integrated QD emission as a function of laser wavelength for the case of resonant cavity excitation [as in (b)]. The solid line is a Lorentzian fit (with a linewidth of 0.1517 nm).

TABLE I: Details of the QD-cavity systems employed in the first experiment, when the cavity emission is observed by resonantly exciting the QD. Also shown are the fits for two different contributions to the QD linewidth, $\Delta\omega_c$ and $\Delta\omega_0$, and the theoretical estimate for $\Delta\omega_c$ (see Eq. 10).

| QD | Temperature (K) | QD Wavelength (nm) | Cavity Wavelength (nm) | $\Delta\omega_c/2\pi$ (Fit) (GHz) | $\Delta\omega_0/2\pi$ (Fit) (GHz) | $\Delta\omega_c/2\pi$ (Theory) GHz |
|---|---|---|---|---|---|---|
| $S1$ | 32 | 934.15 | 934.8 | 12.6 | 1.96 | 1.3 |
| $S2$ | 44 | 932.3 | 931.9 | 9.9 | 9.8 | 2.34 |
| $S3$ | 55 | 933.15 | 931.2 | 15 | 5.8 | 0.28 |

In the first experiment, we observe saturation of the cavity emission with increasing power of the laser used to excite the QD. We fit the cavity intensity with the model given by Eq. 8 [Figs. 3 (a),(c), and (e) (solid line)]. In actual experiments, $\Omega^2 \propto \eta P$, where $P$ is the measured laser excitation power in front of the objective lens and $\eta$ is a constant factor signifying the percentage of incident light coupled to the QD. Hence, assuming that both the QD spontaneous emission rate $2\gamma$ and the pure dephasing rate $\gamma_d$ are independent of the laser excitation power, $\tilde{P} = \alpha P$, where $\alpha$ is a constant factor, independent of the laser power. $\alpha$ is determined from the fit to the cavity intensity with the excitation laser power. In addition to emission saturation, we see broadening of the QD linewidth with increasing excitation laser power, as measured from Lorentzian fits similar to the one shown in Fig. 2 (c). Measurements of the QD linewidth as a function of the laser power for the three different QDs studied are plotted in Figs. 3 (b), (d), and (f). Using the extracted values of $\tilde{P} = \alpha P$ (as previously explained), the linewidths are fit with the model given by Eq. 10 [Figs. 3 (b),(d), and (f) (solid line)]. The fitting parameters are shown in Table I.

We note that for the QD $S1$, the value of $\Delta\omega_0$ obtained from the fit is of the same order of magnitude as the linewidth of a resonantly driven QD without a cavity ($\Delta\omega/2\pi \sim 2.5$ GHz) [14], although in this case we use an off-resonant cavity for read-out. Relatively higher values of $\Delta\omega_0$ for the second ($S2$) and the third ($S3$) QD can be attributed to high dephasing rate at higher sample temperature [16] and the vicinity of etched surfaces of the photonic crystal.

To theoretically estimate $\Delta\omega_c/2\pi$ (contribution from the increased emission into the cavity mode as given by Eq. 4) in Table I , we assume $g = \kappa$. This is an overestimated value of $g$ as our system is not strongly coupled (which is confirmed by bringing the QD onto resonance with the cavity). The overestimated $g$ leads to an overestimate of $\Delta\omega_c$. However, we find that even those theoretically overestimated $\Delta\omega_c$ values are still much lower than the experimental data shown in Table I. Just pure QD dephasing cannot explain this finding as dephasing contributes only to the term $\Delta\omega_0$. The increased broadening indicates a higher coupling strength between the QD and the cavity exceeding what our theoretical model predicts. One possible explanation of this incoherent coupling is that the resonantly excited QD couples to the continuum states provided by the wetting layer or neighboring GaAs layers via tunneling [17] or Auger process [18]. This continuum of states then couples to the off-resonant cavity leading to the observation of cavity emission.

We now analyze the linewidth of the process [Fig. 2(d)] responsible for transferring photons from the resonantly excited cavity to the QD. We perform the second type of experiment (exciting the cavity and collecting emission from the QD) on two QD-cavity systems (Table II). The QD described in the first row of Table II is the same as the QD used in the first experiment (second row of Table I). The other two systems shown in Table I could not be employed in this experiment, as they either showed no emission or very weak emission from QD line under cavity excitation. Hence, we employed another QD system ($S4$)

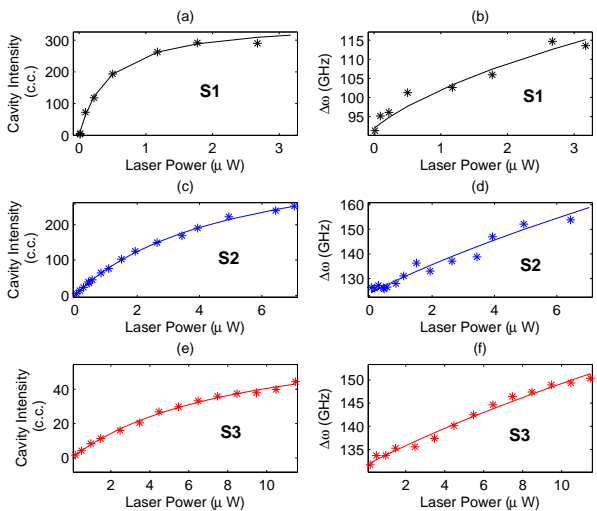

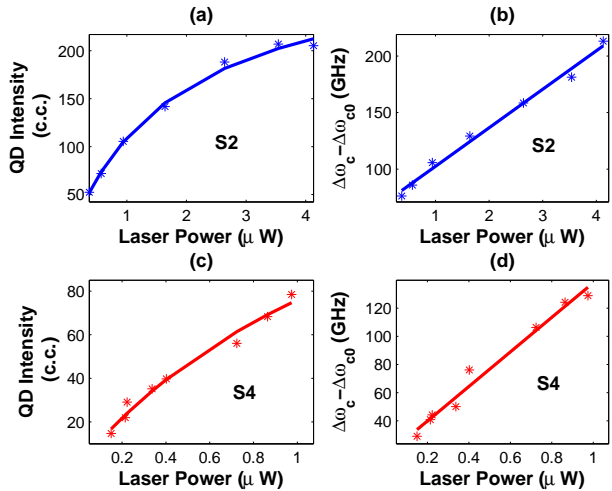

FIG. 3: (color online) (a),(c),(e): Integrated cavity emission as a function of the excitation power of the laser resonantly pumping the QD, for the three QD-cavity systems studied. (c.c. stands for CCD count.) The solid lines are fits to the data using the model given by Eq. 8. (b),(d),(f): Corresponding measured linewidths [as in Fig. 2 (c)] as a function of the laser excitation power. The solid lines are fits to the data using the model given by Eq. 10. The excitation laser power is measured in front of the objective lens.

TABLE II: Details of the QD-cavity systems employed in the second experiment, when the QD emission is observed by resonantly exciting the cavity. Also shown are the values of the $\Delta\omega_{c0}$.

| QD | Temperature (K) | QD Resonance (nm) | Cavity Resonance (nm) | $\Delta\omega_{c0}/2\pi$ (GHz) |
|---|---|---|---|---|
| $S2$ | 44 | 932.3 | 931.9 | 35.6 |
| $S4$ | 55 | 931.9 | 931.2 | 50.3 |

described in Table II.

Figs. 4 (a),(c) show the QD intensity as a function of the power of the laser resonantly pumping the cavity. We observe saturation of the integrated QD emission and the data fit well with the model given by Eq. 8. In this experiment, we also measure the cavity linewidth $\Delta\omega_c$, but here we scan the laser wavelength across the cavity and collect the integrated emission from the QD. In addition, we also measure the intrinsic cavity linewidth $\Delta\omega_{c0}$ from cavity reflectivity measurements at low laser power. In reflectivity measurements, the laser is scanned across the cavity linewidth and the cavity reflected laser power is observed, as in our previous work [1]. For both cavities, the linewidths $\Delta\omega_c$, extracted from the second type of experiment (exciting cavity resonantly and imaging emission at QD wavelength) are larger than the linewidth $\Delta\omega_{c0}$ obtained in reflectivity measurements. Figs. 4 (b),(d) show the difference between two linewidths, i.e.,

FIG. 4: (color online) (a),(c): Integrated QD emission as a function of the excitation power of the laser resonantly pumping the cavity, for the two QD-cavity systems studied (see Table II). (c.c. stands for CCD count.) The solid lines are fits to the data using model given by Eq. 8. (b),(d): The difference between the cavity linewidth $\Delta\omega_c$ measured by observing the QD emission and the cavity linewidth $\Delta\omega_{c0}$ obtained from the cavity reflectivity measurements, as a function of the laser power. The solid line is a linear fit to the difference. The excitation laser power is measured in front of the objective lens.

$(\Delta\omega_c - \Delta\omega_{c0})$, which increases linearly with laser power. This additional broadening is attributed to the free carriers generated by the laser excitation.

In conclusion, we studied the off-resonant QD-cavity coupling under resonant excitation of both the QD and the cavity. We found that pure dephasing along with power broadening and coherent coupling between the cavity and the QD underestimate the QD linewidth. This indicates a higher incoherent coupling strength between the QD and the cavity, possibly resulting from the coupling to the continuum of states of the wetting layer or neighboring GaAs [17, 18].

The authors acknowledge financial support provided by the National Science Foundation, Army Research Office and Office of Naval Research. A.M. was supported by the Stanford Graduate Fellowship (Texas Instruments fellowship).